\documentclass[pra,twocolumn,superscriptaddress,bibnotes]{revtex4-1}

\usepackage[a4paper,centering,hmargin=1.75cm,vmargin=2cm]{geometry}
\usepackage{amssymb,amsmath,amsfonts}
\usepackage{microtype,graphicx,xcolor,braket,siunitx}
\usepackage[colorlinks,allcolors=blue!50!black]{hyperref}
\usepackage[all]{hypcap}

\begin{document}

\title{Geometry, supertransfer,  and optimality in the light harvesting of purple bacteria}

\author{Sima Baghbanzadeh}
\affiliation{Department of Physics, Sharif University of Technology, Tehran 11155-9161, Iran}
\affiliation{Centre for Engineered Quantum Systems and School of Mathematics and Physics, The University of Queensland, Brisbane Queensland 4072, Australia}
\affiliation{School of Physics, Institute for Research in Fundamental Sciences (IPM), Tehran 19395-5531, Iran}
\author{Ivan Kassal}
\email{Email: i.kassal@uq.edu.au}
\affiliation{Centre for Engineered Quantum Systems and School of Mathematics and Physics, The University of Queensland, Brisbane Queensland 4072, Australia}

\begin{abstract}
The remarkable rotational symmetry of the photosynthetic antenna complexes of purple bacteria has long been thought to enhance their light harvesting and excitation energy transport. We study the role of symmetry by modeling hypothetical antennas whose symmetry is broken by altering the orientations of the bacteriochlorophyll pigments. We find that in both LH2 and LH1 complexes, symmetry increases energy transfer rates by enabling the cooperative, coherent process of supertransfer. The enhancement is particularly pronounced in the LH1 complex, whose natural geometry outperforms the average randomized geometry by 5.5 standard deviations, the most significant coherence-related enhancement found in a photosynthetic complex.
\end{abstract}

\maketitle

Photosynthetic organisms use light-harvesting antenna complexes to absorb light and funnel the resulting excitation energy into a reaction center (RC), where the energy is used to drive charge separation~\cite{Blankenship2014}. Despite the diversity of antenna complexes, the efficiency of excitation energy transfer (EET~\cite{MayKuhn}) through them is generally high, prompting hopes that understanding EET mechanisms in these complexes will generate new ideas for improving artificial light harvesting~\cite{Scholes:2011uj,Laos:2014jf}.

In searching for design principles in photosynthetic architectures, it is important to not assume that a particular photosynthetic system is optimized simply because it is a product of billions of years of natural selection. If nothing else, the dramatically different antenna architectures in different plant and bacterial taxa~\cite{Blankenship2014} cannot all be optimal. In other words, the optimality of photosynthetic light harvesting is a hypothesis to be tested, with there being a distinct possibility that a particular arrangement is not optimal but is merely good enough to ensure the particular organism's survival.

A way to determine whether an EET architecture is optimal is to examine its performance if its structure is changed in significant ways~\cite{Cao2009,Scholak2011,Mohseni:2013ik,Wu2013}. This kind of analysis has been carried out for the photosynthetic apparatus of several species. For example, in a model of the cyanobacterial photosystem~I~(PSI), randomizing the orientations of the chlorophyll (Chl) molecules about their Mg atoms altered the overall quantum yield by less than 1\%, and, indeed, the already high yield could be further increased by adding small variations in site energies~\cite{MSener2002}. Similarly, in a kinetic model of photosystem~II (PSII), randomizing the Chl orientations changed the yield by up to a few percent, with the X-ray geometry being near the middle of the distribution~\cite{Vasileva2004}. To us, these findings suggest that Chl orientations in neither PSI nor PSII are fine-tuned to an optimal geometry, especially considering that the uncertainties in the approximations employed exceeded the maximum claimed few-percent enhancement. 

EET through the Fenna-Matthews-Olson (FMO) complex of green sulfur bacteria has also been widely studied, and it has been claimed to be close to optimal with respect to not only geometric changes but also changes in environmental parameters such as the temperature or the reorganization energy~\cite{Mohseni2008,Rebentrost2009,Plenio2008,Chin2010,Wu2010,Mohseni:2014iv,Shabani:2014tt,Dubi2015}. However, the optimality of EET in FMO is not settled, since treating photoexcitation realistically is likely to substantially affect the efficiency~\cite{LeonMontiel2014}.

\begin{figure}[b]
    \centering
    \includegraphics[width=0.9\linewidth]{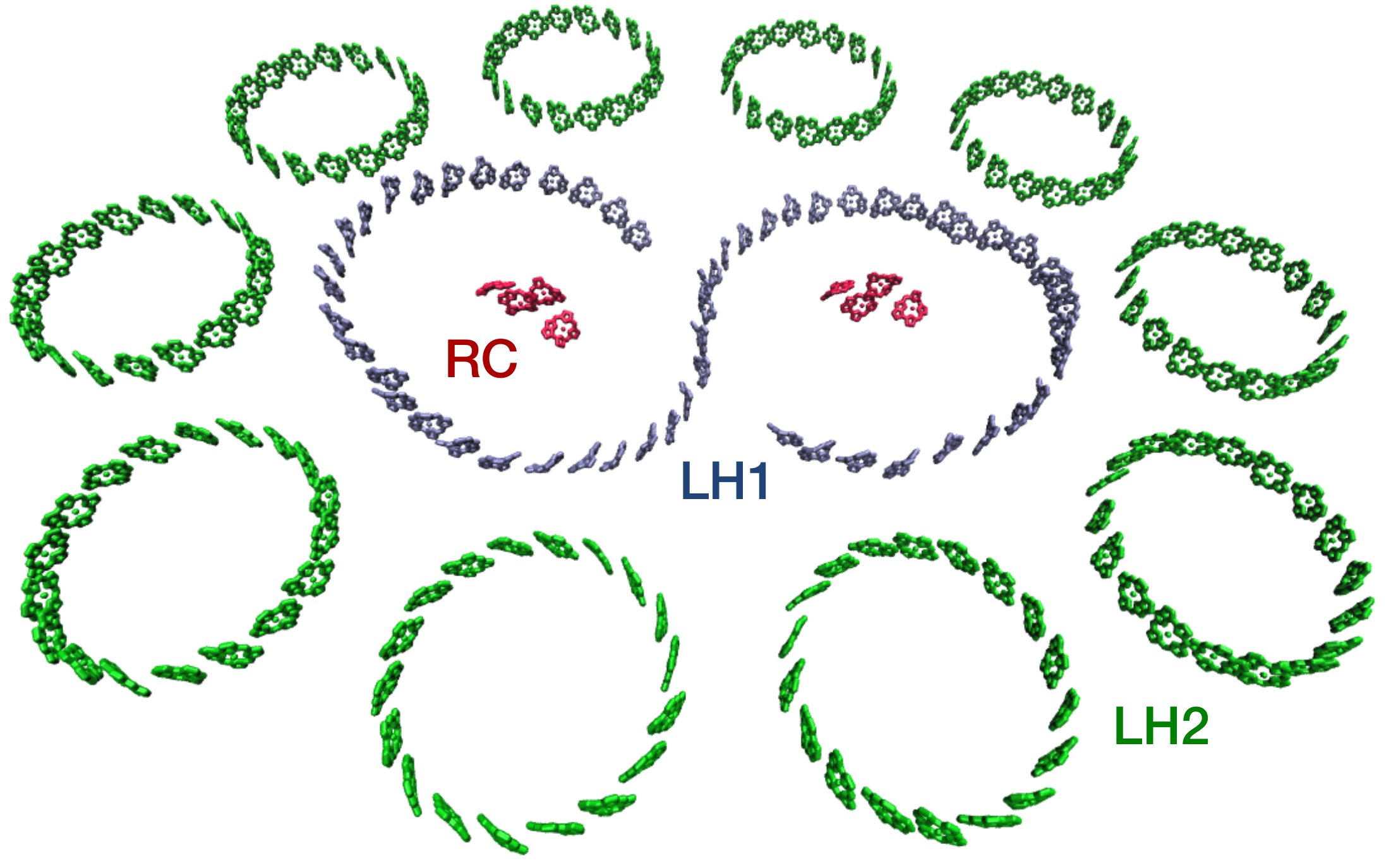}
    \caption{Model of the photosynthetic apparatus of \textit{Rh.\;sphaeroides}, including the reaction centre (RC) surrounded by the antenna complexes LH1 and LH2 (only B850 subunit shown). Reproduced with permission from~\cite{Baghbanzadeh2015}. Copyright 2016 Royal Society of Chemistry.}
    \label{fig:structure}
\end{figure}

Here, we consider the light-harvesting apparatus of purple bacteria, well known for their highly symmetric antenna complexes LH2 and LH1~\cite{Cogdell:2006ko}, shown in Fig.~\ref{fig:structure}. EET through these complexes has been studied extensively~\cite{Hu1997,Scholes:1999tm,Damjanovic2000,Ritz2001,Sener2007,Olaya-Castro2008,Linnanto2009,Strumpfer2009,Sener:2010ct,Strumpfer2012a,Strumpfer2012b,Cleary:2013dl}, often using kinetic models. Most models recognize the importance of the strong couplings between bacteriochlorophylls (BChls) in LH2 and LH1, which gives rise to considerable excitonic delocalization~\cite{Leupold1996,Pullerits1996,Monshouwer1997,Oijen1999}. Delocalization is particularly relevant in LH2 and LH1, where it can lead to supertransfer~\cite{Strek1977,Mukai1999,Scholes2000,Scholes:2002ie,Jang2004,Jang2007,Lloyd:2010fz,Strumpfer2012b}, an enhancement of EET rates over site-to-site hopping because the dipole moments of individual pigments are oscillating with a definite phase. 

Limited work has been done on whether the efficiency of purple-bacterial light harvesting is optimal with respect to certain parameters. In particular, it has been suggested that the internal symmetry of LH2 is particularly well-suited to maximizing the packing density and minimizing frustration within a larger lattice~\cite{Cleary:2013dl}. In addition, we have previously considered the efficiency of purple-bacterial light harvesting after changes to site energies and after suppressing delocalization by trimming away every second BChl~\cite{Baghbanzadeh2015}. Trimming frequently reduced the efficiency by a large margin, confirming the influence of delocalization on light harvesting. However, we also showed that delocalization is not necessary for high efficiency, since a decrease in performance could always be compensated by altering the site energies to create an energy funnel into the RC.

Here, we investigate the efficiency of purple-bacterial light-harvesting in natural light conditions as the BChl orientations are changed, finding that the performance of the natural geometry is one of the highest among thousands of reorientations. There is also a marked difference between the robustness of LH2 and LH1 to pigment reorientation. Whereas changes to LH2 hardly affect the efficiency, the natural orientations in LH1 are a significant outlier, lying 5.5 standard deviations from the mean. We attribute this sensitivity to supertransfer, which reaches its maximum near the natural orientations. Indeed, because supertransfer is a consequence of excitonic delocalization, we show that the efficiency is less sensitive to geometric effects if delocalization is turned off by removing every second pigment.

\section{Model}

We consider the photosynthetic apparatus of the purple bacterium \textit{Rhodobacter sphaeroides} using the model described previously~\cite{Baghbanzadeh2015}. As shown in Fig.~\ref{fig:structure}, it includes antenna complexes LH1 and LH2 that increase the amount of light absorbed per RC. Two RCs are surrounded by the S-shaped LH1 complex consisting of $56$ tightly packed BChls, which is itself surrounded by LH2 complexes, the structures being taken from crystal structures~\cite{Papiz2003,Qian2013}. Although each LH2 contains two rings of BChls, B800 and B850, we only consider the 18-member B850 because EET between B800 and B850 is fast and efficient. Overall, the main inter-complex EET pathway is LH2~$\to$~LH1~$\to$~RC.

LH2 has a beautiful 9-fold symmetry, with the transition dipoles of the B850 BChls almost in the plane of the ring (about \SI{5}{\degree} off), pointing alternately left and right as they go around. Although LH1 is less symmetric than LH2, its BChls are approximately arranged on a flat ring within each of its halves, with their transition dipole moments again roughly parallel to the plane of the ring (at most \SI{25}{\degree} off) and also alternating left-right.

Strong coupling between nearest-neighbor BChls results in exciton delocalization within each complex, i.e., LH2, LH1, or RC. The excitonic states are eigenstates of a Frenkel-type Hamiltonian of the particular complex~\cite{MayKuhn}, which---in weak light where there is at most one exciton present---takes the form
\begin{eqnarray}
H=\sum_i^N E_i |i\rangle\langle i| + \sum_{i<j}^N V_{ij}(|i\rangle\langle j|+|j\rangle\langle i|),
\end{eqnarray}
where $N$ is the number of pigments within the complex, $E_i$ is the ``site'' energy of state $|i\rangle$ corresponding to an exciton on BChl $i$, and $V_{i,j}$ is the coupling between sites $i$ and $j$. 

Different computational methods can predict substantially different energies and couplings~\cite{Hu1997,Koolhaas1998,Scholes:1999tm,Tretiak2000-2,Linnanto2009,Baghbanzadeh2015}. Here, we compute the intra- and inter-complex couplings $V_{i,j}$ using Transition charges from electrostatic potentials (TrEsp)~\cite{Madjet2006}, a method that is not only fast, but also as accurate as is realistically possible across the full range of (bacterio)chlorophyll separations and orientations~\cite{Kenny2015}. The only exception is the RC special pair, whose coupling we take to be $\SI{418}{cm}^{-1}$~\cite{Madjet2009}. Furthermore, for each complex, we choose the site energies so that the energy of the brightest state matches the observed absorption maximum of that complex.

Because couplings between different complexes are weak, we neglect inter-complex excitonic delocalization. Accordingly, optical pumping and dynamics will be entirely through the eigenstates of the different complexes, as opposed to individual sites~\cite{MayKuhn,Mancal2010,Kassal2013,LeonMontiel2014}. In particular, EET between two weakly coupled aggregates is described by multichromophoric F\"orster resonant energy transfer (MC-FRET)~\cite{Sumi1999,Jang2004,Jang2007}. MC-FRET simplifies to the more tractable generalized F\"orster resonant energy transfer (gFRET)~\cite{Mukai1999,Scholes2000,Scholes2003} in several cases, including if the emission and absorption spectra of the complexes are diagonal in the excitonic basis or if the system-environment coupling is weak compared to the coupling between BChls in the same complex~\cite{Jang2007}. Assuming the latter, the gFRET transfer rate between eigenstates of two complexes is 
\begin{equation}
k_{\phi\psi}^{\mathrm{ET}}=\frac{2\pi}{\hbar}|V_{\phi\psi}|^2 J_{\phi\psi},
\label{eq:FRETrate}
\end{equation}
where $V_{\phi\psi}=\sum_{i,j} c^{\psi}_i c^{\phi}_j V_{ij}$, $c^{\psi}_i$ and $c^{\phi}_j$ are the components of the excitonic states $\psi$ and $\phi$ in the site basis, and $V_{ij}$ is the coupling between sites $i$ and $j$. $J_{\phi\psi}=\int L_\psi(E)I_\phi(E)\,dE$ is the spectral overlap between the normalized emission spectrum $L_\psi$ of the donor and the normalized absorption spectrum $I_\phi$ of the acceptor. $L_\psi$ and $I_\phi$ can be calculated using multichromophoric FRET theory~\cite{Jang2004,Ma:2015kg,Ma:2015ju,Moix:2015ei}, but here we follow~\cite{Baghbanzadeh2015} in taking both to be normalized Gaussians, giving $J_{\phi\psi} = \exp(-E_{\phi\psi}^2/4\sigma^2)/\sqrt{4\pi\sigma^2}$, where $E_{\phi\psi}$ is the energy difference between the states and $\sigma=\SI{250}{cm^{-1}}$. Finally, to ensure detailed balance, we use Eq.~\ref{eq:FRETrate} only for downhill transitions ($E_\psi > E_\phi$), otherwise taking $k_{\phi\psi}^{\mathrm{ET}} = k_{\psi\phi}^{\mathrm{ET}} e^{-E_{\phi\psi}/k_\mathrm{B}T_\mathrm{B}}$ with $T_\mathrm{B}=\SI{300}{K}$.

Because the site-to-site couplings in the generalized FRET expression (Eq.~\ref{eq:FRETrate}) are combined with amplitudes $c^{\psi}_i$ and $c^{\phi}_j$ that can be positive or negative (or complex), the overall rate $k_{\phi\psi}^{\mathrm{ET}}$ can be larger or smaller than an analogous incoherent sum of site-to-site FRET rates. When the excitonic states are delocalized so that the amplitudes cause a cooperative enhancement of the transfer rate, the effect is called supertransfer~\cite{Strek1977,Scholes:2002ie,Lloyd:2010fz}.

\begin{figure*}[t]
    \centering
    \includegraphics[width=\textwidth]{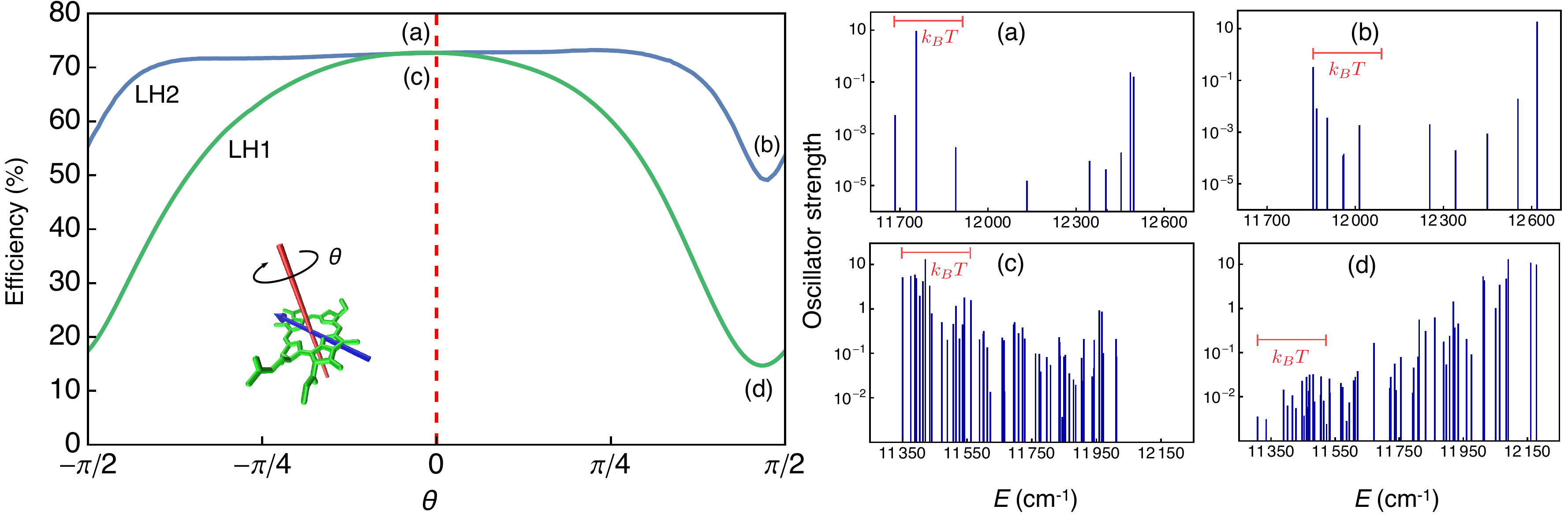}
    \caption{Light-harvesting efficiency when the BChls within either LH2 (blue) or LH1 (green) are rotated in the plane of their bacteriochlorin rings by angle $\theta$ (see \textbf{Inset}, which also shows the direction of the transition dipole moment as the blue arrow). The natural geometry ($\theta=0$) is close to optimal, with substantial decreases in efficiency as $\theta$ changes. The high efficiency at $\theta=0$ is caused by the very bright low-lying (thermally accessible at $k_\mathrm{B}T=\SI{200}{cm^{-1}}$) states of both LH2~\textbf{(a)} and LH1~\textbf{(c)}, which encourages the onward supertransfer of excitons. The efficiency reaches a minimum close to $\theta=\pi/2$ (transition dipoles perpendicular to the plane of the rings) for both LH2~\textbf{(b)} and LH1~\textbf{(d)} because the thermally accessible states are now much darker. Roughly speaking, the complex changes from a $J$~aggregate to an $H$~aggregate~\cite{MayKuhn} as $\theta$ changes. Note that the asymmetry of the curves around $\theta=0$ is due to the fact that the dipole moments of the pigments in the complexes are not exactly in the plane of the relevant rings. On the average, dipole moments within LH2 and LH1 are, respectively, \SI{5}{\degree} and \SI{7}{\degree} out of the plane.}
    \label{fig:in-plane}
\end{figure*}

In natural light, optical pumping and relaxation take place continuously, meaning that the ensemble of complexes will reach a steady state, finding which is sufficient to determine all observables of interest~\cite{Kassal2013,Baghbanzadeh2015}. Because sunlight is incoherent, it creates excitons in energy eigenstates~\cite{Jiang:1991fk,Brumer:2011ty,Mancal2010}, i.e., it does not induce coherences in the energy basis. Strictly speaking, the incoherent pumping is into eigenstates of the combined system and bath, which, when reduced to the system alone, may not coincide with eigenstates of the system Hamiltonian~\cite{Olsina:2014vb,Tscherbul2014,Tscherbul2015}. Here we assume that the system-bath coupling is not large enough for this discrepancy to be significant. 

Consequently, the dynamics of the apparatus can be described using a Pauli master equation, $\dot{\mathbf{p}} = K \mathbf{p}$, where $\mathbf{p}$ is the vector of all eigenstate populations, including the ground state. The rate matrix $K$ contains the absorption, relaxation, and inter-complex transfer rates,
\begin{eqnarray}
 K_{\phi\psi} &=& k_{\phi\psi}^{\mathrm{ET}} + k_{\phi\psi}^{\mathrm{RR}} + k_{\phi\psi}^{\mathrm{NR}} + k_{\phi\psi}^{\mathrm{IC}} + k_{\phi\psi}^{\mathrm{OP}} + k_{\phi\psi}^{\mathrm{CS}} \nonumber \\
		&&   (\text{for}\;\; \phi\ne\psi), \nonumber\\
 K_{\phi\phi} &=& -\! \sum_{\phi\ne\psi} K_{\psi\phi}.
\end{eqnarray}
Here, nonradiative recombination to the ground state $g$ is assumed to occur at rate $k_{g\phi}^\mathrm{NR}=(\SI{1}{ns})^{-1}$~\cite{Sener2007} and internal conversion to lower-lying excitonic levels with rate $k_{\phi\psi}^\mathrm{IC}=(\SI{100}{fs})^{-1}$~\cite{Baghbanzadeh2015}. Radiative recombination is taken to occur with rate $k^\mathrm{RR}_{g\phi}=k^\mathrm{RR}_{0} f_\phi  (E_{\phi}/E_0)^3$ where $k^\mathrm{RR}_{0}=(\SI{16.6}{ns})^{-1}$~\cite{Monshouwer1997} and $E_0=hc/(\SI{770}{nm})$ are, respectively, the radiative decay rate and site energy of BChl in solution, while $f_\phi=|\mu_{\phi}/\mu_0|^2$ is the oscillator strength (or brightness) of state $\phi$ relative to a single BChl. The optical pumping rate is $k^\mathrm{OP}_{\phi g}=k^\mathrm{RR}_{g\phi}\,n(E_\phi)$, where $n(E_\phi)=(e^{E_\phi/k_\mathrm{B}T_\mathrm{R}}-1)^{-1}$ is the mean photon number at energy $E_\phi$ at the effective black-body temperature of solar radiation, $T_\mathrm{R}=\SI{5780}{K}$. Finally, $k^\mathrm{CS}_{g\phi}=(\SI{3}{ps})^{-1}$ (if $\phi$ is a state of the RC) is the rate of charge separation in the RC~\cite{Blankenship2014,Sener2007}.

At steady state $\dot{\mathbf{p}}^\mathrm{SS}=0$ and the problem simplifies to finding the zero-eigenvalue eigenstate of $K$, whose existence and uniqueness are guaranteed because $K$ describes an irreducible continuous-time Markov chain. The overall EET efficiency, 
\begin{equation}
\eta = \frac{k^\mathrm{CS} \sum_{\phi\in \mathrm{RC}} \, p_\phi^\mathrm{SS}}{p_g^\mathrm{SS}\,\sum_{\phi} k^\mathrm{OP}_{\phi g}},
\label{eq:efficiency}
\end{equation}
is the quantum yield of photoexcited excitons that drive charge separation in the RC at steady state.

\section{Results and Discussion}

We study how the orientations of BChls within LH2 and LH1 aggregates affect the efficiency of exciton transfer in the purple-bacterial light-harvesting apparatus. In doing so, we fix the site energies---whose role we examined previously~\cite{Baghbanzadeh2015}---and the positions of the central Mg atom in each BChl. Rotating the BChls has a complex influence on the performance of the complexes because changing the orientations of the dipole moments affects inter-pigment couplings and thus the nature and energy of the eigenstates. These changes, in turn, affect the inter-complex EET rates and the overall efficiency.

We consider two types of changes: rotating BChls within their planes as well as randomizing their orientations completely.

\begin{figure*}[t]
    \centering
     \includegraphics[width=\textwidth]{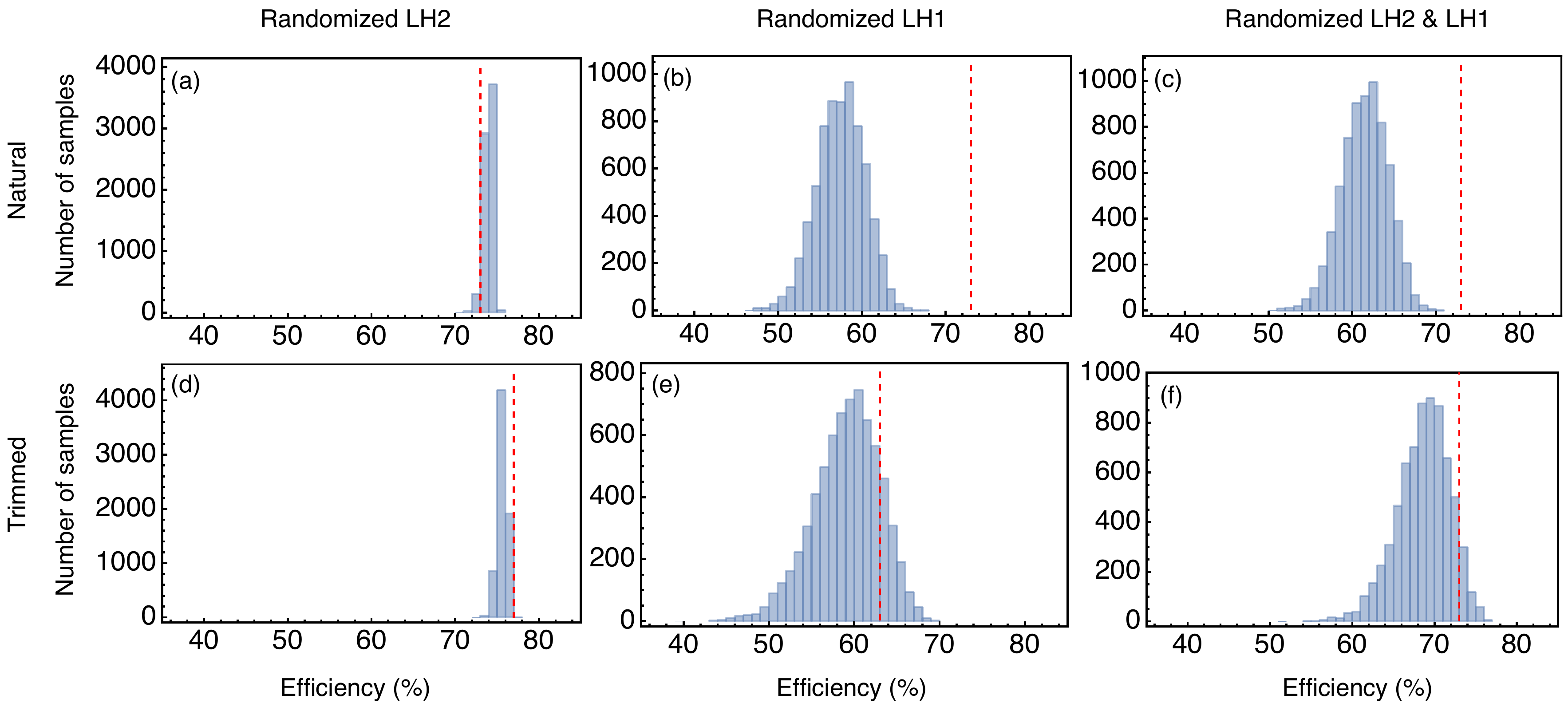}
    \caption{Distribution of the efficiency as the orientations of the BChls are completely randomized in the LH2 complexes \textbf{(a)}, in LH1 \textbf{(b)}, or in both~\textbf{(c)}, with $7000$ realizations in each case. In each panel, the dashed red line indicates the efficiency at the natural geometry, showing that reorientations in LH2 do not affect the efficiency, while those in LH1 reduce it significantly. Indeed, efficiency when LH1 is at its natural geometry is an outlier by 5.5 standard deviations. 
    \textbf{(d--f)} Same as panels (a--c), except that the complexes are trimmed by removing every second BChl, suppressing excitonic delocalization. In particular, the natural LH1 geometry is no longer an outlier, indicating that delocalization enhances the natural efficiency through supertransfer.
    }
    \label{fig:EfficiencyDistributions}
\end{figure*}

\subsection{In-plane rotations of BChls}

The bacteriochlorin ring within each BChl is approximately planar, meaning that it would occupy roughly the same space if it were rotated about an axis passing through the Mg atom and perpendicular to the ring (see inset to Fig.~\ref{fig:in-plane}). Thus, in-plane BChl rotations might be considered more plausible evolutionary alternatives than some other rotations, since the bacteriochlorin ring would not require large adjustments in the surrounding protein. Although this argument neglects the BChl’s phytyl tail, the tail is of secondary importance because it would be flexible enough to bend out of the way in many cases of steric hindrance. Most importantly, the simple rotation provides substantial intuition about the role of BChl orientations that can be used to understand more complicated rearrangements.

Figure~\ref{fig:in-plane} shows how in-plane BChl rotations affect the overall efficiency. Two cases are shown, with all the BChls in either LH2 or LH1 rotated by the same angle $\theta$. The X-ray geometry corresponds to $\theta=0$, whose efficiency (73\%) is nearly optimal. Indeed, rotating the BChls can reduce the efficiency significantly, as low as 15\% in the case where LH1 BChls are set to be perpendicular to the LH1 plane. 

The reduction in efficiency upon BChl rotation can be understood by considering the brightnesses of the aggregate energy levels. If the aggregates were far apart compared to their size, each could be considered as a supermolecule, with FRET rates proportional to the oscillator strengths of the excitonic states. Here, the small inter-complex distances mean that the supermolecule approximation does not capture all the details of Eq.~\ref{eq:FRETrate}, but it is nevertheless a useful conceptual tool. We stress that the brightness of the states relates to the efficiency because it is a proxy for supertransfer, not because it implies that more light is absorbed in the first place. On the contrary, the oscillator sum rule implies that the total absorption cross-section of all the states will be constant regardless of their individual brightnesses, assuming the solar intensity is approximately constant over the absorption spectrum.
 
As depicted in Fig.~\ref{fig:in-plane}(a) and (c), the natural geometries of both LH2 and LH1 give rise to very bright states near the bottom of the energy spectrum (within $k_\mathrm{B}T=\SI{200}{cm^{-1}}$ of the lowest state). This is important because the high internal-conversion rate $k^\mathrm{IC}$ ensures rapid thermalization, meaning that only the low-lying states contribute to EET and that their brightness regulates supertransfer. As the dipole moments rotate away from the plane of the rings, the low-lying bright states are gradually lost until, at the minima shown in Fig.~\ref{fig:in-plane}(b) and (d), the thermally accessible states carry very little oscillator strength. Thus, EET is slowed down and efficiency decreases.

Figure~\ref{fig:in-plane} shows that the efficiency is more sensitive to the reorientation of BChls in LH1 than in LH2. This is because LH1~$\to$~RC transfer is the kinetic bottleneck of the entire process, largely because it is energetically uphill (for the complete energy diagrams of LH2, LH1, and RC, see ref.~\cite{Baghbanzadeh2015}). Therefore, decreases in the LH1~$\to$~RC EET rate caused by reorientation immediately translate to a reduced efficiency. By contrast, LH2~$\to$~LH1 transfer is energetically downhill and relatively fast, meaning that it can proceed with high efficiency even if the rate decreases somewhat. Thus, the broad plateau in the efficiency as a function of LH2 rotation angle reflects the need for a significant rotation of the BChls before the rate is decreased enough for it to affect the overall efficiency. Even at the minimum, the efficiency only decreases from 73\% to 49\%, reflecting the decisive effect of downhill energetic funnelling~\cite{Baghbanzadeh2015}. 

\begin{figure*}[t]
    \centering
     \includegraphics[width=0.8\textwidth]{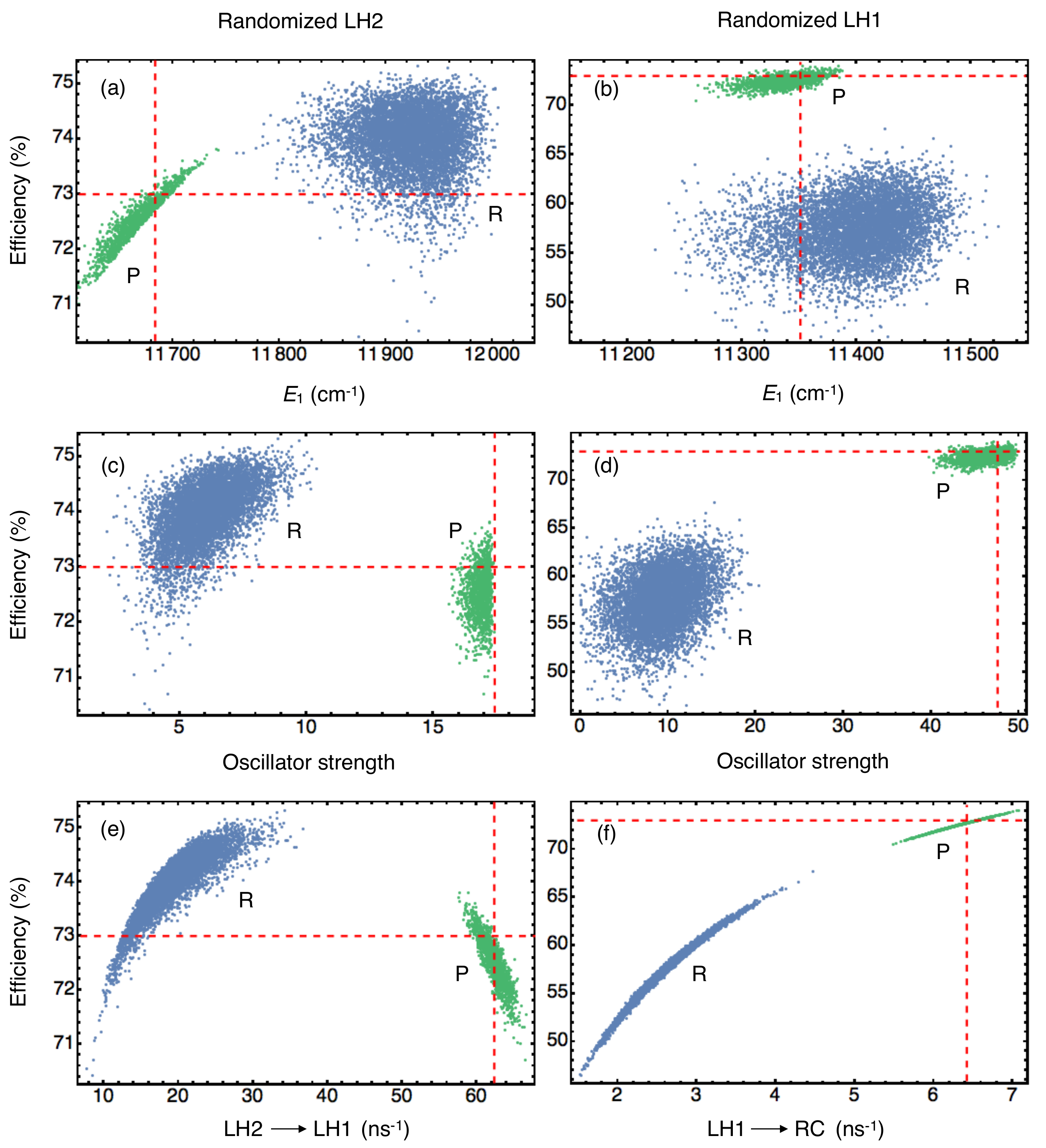}
    \caption{Determining why the efficiency is more sensitive to geometric changes in LH1 than in LH2. In each panel, the completely randomized orientations are denoted with blue dots (R) and the perturbed orientations (within \SI{5}{\degree} for each BChl) with green dots (P). The dashed red lines indicate parameter values at the natural geometry.
   \textbf{(a)} In LH2, randomization increases the energy $E_1$ of the lowest excitonic state, but with no significant effect on the efficiency.
   \textbf{(b)} In LH1 as well, there is no correlation between excitonic energies and the efficiency.
    \textbf{(c)} In LH2, randomization decreases the total brightness of the low-lying states (those within $k_\mathrm{B} T$ of $E_1$) approximately three-fold, with no significant decrease in efficiency.
    \textbf{(d)} In LH1, the roughly four-fold decrease in brightness \emph{does} lead to a large decrease in efficiency.
    \textbf{(e)} In LH2, the reduced brightness is reflected in the reduced energy transfer rate to LH1, but the rate is high enough that the reduction does not affect the overall efficiency.
    \textbf{(f)} In LH1, the transfer rate to the reaction centre is the kinetic bottleneck, and even in the natural geometry the rate is low enough to become comparable to exciton loss through recombination (at a rate of \SI{1}{ns^{-1}}). Slowing this process down causes the decrease in efficiency. 
    }
    \label{fig:ScatterPlots}
\end{figure*}

\subsection{Complete randomization of BChl orientations}

To further investigate the influence of geometry on EET efficiency in purple bacteria, we considered aggregates in which the orientations of the BChls in LH2 and/or LH1 were completely randomized, the orientations being chosen using a standard spherical point-picking algorithm.  Because random rotations could cause nearest-neighbor BChls to collide with each other, we only accepted geometries in which the distance between any two atoms in different BChls is greater than $\SI{2.36}{\AA}$, which is the shortest distance between BChls in LH1 aggregate and is approximately twice the van der Waals radius of a hydrogen atom.

The distributions of efficiencies for the random orientations are shown in Fig.~\ref{fig:EfficiencyDistributions}(a--c). In particular, the efficiency is not sensitive to the orientation of BChls in the LH2 complexes, always attaining a value close to the original $73\%$ (Fig.~\ref{fig:EfficiencyDistributions}(a)). This indicates that no geometric fine tuning is necessary to achieve a high efficiency and that LH2s are tolerant to orientational disorder.

By contrast, BChl orientations in LH1 have a large effect on the efficiency (Fig.~\ref{fig:EfficiencyDistributions}(b)). Importantly, the mean efficiency is $57\%$, with none of the $7000$ samples coming close to the original $73\%$, making the natural geometry an outlier by 5.5 standard deviations. It follows that the whole light-harvesting apparatus has an unusually high efficiency, as is seen when the BChls in both LH2 and LH1 are randomized (Fig.~\ref{fig:EfficiencyDistributions}(c)). 

The natural LH1 geometry is an outlier because it occupies a corner of an enormous, 168-dimensional space (three angles per BChl). If only small perturbations to the original BChl angles were considered (up to~\SI{5}{\degree}), the efficiency would only change by up to a few percent (see Fig.~\ref{fig:ScatterPlots}). Indeed, it is unlikely that BChl orientations are fine-tuned to less than several degrees, considering the constant fluctuations at physiological temperatures.

The stark difference between the effect of randomization on LH2 and LH1 can be understood, as in the case of in-plane rotations, in terms of the brightnesses of the states and of the rate-limiting nature of the LH1~$\to$~RC step. To establish this conclusion, Fig.~\ref{fig:ScatterPlots} shows the relationships between the overall efficiency and the two properties that have the greatest influence on EET, energy funnelling and coherent excitonic delocalization~\cite{Baghbanzadeh2015}. Energetic alignment enters $J_{\phi\psi}$ in Eq.~\ref{eq:FRETrate}, strongly favoring downhill EET, while the oscillator strength is a useful, if approximate, proxy for the delocalization and supertransfer contained in $V_{\phi\psi}$.

Fig.~\ref{fig:ScatterPlots}(a) and (b) show there is no appreciable correlation between efficiency and the energy of the lowest excited states in either LH2 or LH1. In LH2, random reorientations increase the energy of the lowest excitonic state a few hundred wavenumbers with no effect on the efficiency because the EET to LH1 remains downhill. In LH1, it might be expected that an increase in energy upon randomization would increase the efficiency by reducing the uphill energy barrier for transfer to the RC. However, the range of energetic variation is comparable to $k_\mathrm{B}T=\SI{200}{cm^{-1}}$, resulting in minor changes to the efficiency compared to other effects.

Fig.~\ref{fig:ScatterPlots}(c) and (d) examine the correlation between efficiency and the brightness of the low-lying states in LH2 and LH1, defined as the sum of oscillator strengths of the states lying within $k_\mathrm{B}T$ of the lowest state. In both LH2 and LH1, randomization destroys the symmetry and decreases the oscillator strength by a factor of 3--4. Nevertheless, there is a large difference between the effect of brightness on efficiency for the two complexes: for LH2 there is no effect, while for LH1 it leads to a large decrease in efficiency. The same difference is seen when considering the rates of forward EET (from LH2 to LH1 and from LH1 to RC): although the decrease in brightness reduces forward EET rates in both LH2 and LH1, only the decrease in LH1 affects the efficiency (Fig.~\ref{fig:ScatterPlots}(e) and (f)). Indeed, in LH1 the decrease in brightness is sufficient to decrease the EET rate despite the improved energetic landscape.

As with in-plane rotations, LH1 is more sensitive to changes in the brightness of its states because the rate of LH1~$\to$~RC transfer is low to begin with; at \SI{6.4}{ns^{-1}}, it is the lowest EET rate in the entire light-harvesting apparatus~\cite{Baghbanzadeh2015} and is comparable to the recombination rate $k^\mathrm{NR}=\SI{1}{ns^{-1}}$. Decreasing it further by reducing brightness tightens the bottleneck, directly resulting in a decrease in efficiency (Fig.~\ref{fig:ScatterPlots}(f)). By contrast, the LH2~$\to$~LH1 rate is high enough even with the decrease in brightness that there is little risk of the exciton being lost while on LH2.

Bright states are a manifestation of excitonic delocalization, and their crucial contribution to the nearly optimal efficiency in the natural geometry can be corroborated by turning delocalization off. Fig.~\ref{fig:EfficiencyDistributions}(d--f) shows the distribution of efficiencies for complexes that are trimmed by removing every second BChl. Doing so doubles the nearest-neighbor distances, weakening intra-complex couplings and suppressing excitonic delocalization, meaning that EET takes place by ordinary, site-to-site FRET~\cite{Baghbanzadeh2015}. In particular, trimming LH1 not only reduces the efficiency from the natural delocalized case, but it also makes it so that the original orientation of BChls is no longer an efficiency outlier. This confirms our claim that the feature which makes the original orientation of the BChls within LH1 an outlier is the coherent excitonic delocalization and the resulting supertransfer. 

\section{Conclusions}

In summary, our study of geometric effects in purple-bacterial energy transfer reveals that, if the site energies are fixed, altering pigment orientations can significantly reduce the efficiency. The effect is due to the fragility of low-lying excitonic states, whose high brightness in the natural geometry yields the high efficiency through supertransfer. The magnitude of the improvement---a natural geometry that is 5.5 standard deviations better than the mean random geometry---is one of the largest photosynthetic efficiency enhancement we are aware of that has been attributed to a coherent effect. 

The natural geometry's exceptional efficiency among plausible evolutionary alternatives suggests that it may have conferred an evolutionary advantage. If so, delocalization would likely be a spandrel, a feature that was originally a byproduct of evolution, but was later exploited to improve fitness~\cite{Gould:1979db,Gould:1997fv}. We argued previously that delocalization was not required for high efficiencies in purple-bacterial light harvesting, suggesting that it arose as a byproduct of the tight bacteriochlorophyll packing that enhances the absorption cross-section per RC~\cite{Baghbanzadeh2015}. But if ring-like structures with delocalization were already present, it is plausible that subsequent evolution adjusted the directions of the dipole moments to take advantage of supertransfer. Of course, these speculations would need to be tested by future work, especially through a comparison of corresponding structures in different taxa of purple bacteria.

More generally, our results confirm the predicted importance of supertransfer as one of the few coherent mechanisms possible in incoherent light~\cite{Kassal2013} and promise to increase its deployment in artificial light-harvesting complexes.

\section{Acknowledgements}

We thank Saleh Rahimi-Keshari for useful discussions. We were supported by the Australian Research Council through a Discovery Early Career Researcher Award (DE140100433) and the Centre of Excellence for Engineered Quantum Systems (CE110001013).

\end{document}